\documentclass[11pt,a4paper]{article}         
\usepackage[utf8]{inputenc}                   
\usepackage[T1]{fontenc}                      
\usepackage{lmodern}                          
\usepackage[english]{babel}                   
\usepackage{amsmath,amsfonts,amssymb}         
\usepackage{amsbsy}
\usepackage{bbold}
\usepackage{gensymb}                          
\usepackage{amsthm}                           
\usepackage{pstricks-add}                     
\usepackage{graphicx}                         
\usepackage{fancyhdr}                         
\usepackage{changepage}
\usepackage{lipsum}
\usepackage{array}
\newcolumntype{P}[1]{>{\centering\arraybackslash}p{#1}}
\usepackage{xcolor}
\usepackage{listings}
\usepackage{multicol}
\usepackage{csquotes}
\usepackage{listingsutf8}
\usepackage{lettrine}
\usepackage{titling}
\usepackage{soul}
\usepackage{todonotes}
\usepackage{authblk}
\definecolor{bleuclair}{cmyk}{0.1,0,0,0.01}
\usepackage[top=3cm, bottom=3cm, left=2cm, right=2cm]{geometry}
\usepackage[title]{appendix}
\usepackage{algorithmic}
\usepackage{fancybox}

\usepackage{hyperref}
	\hypersetup{
	    pdffitwindow=false,               
	    pdfsubject={dimensions.exe},      
	    pdfnewwindow=true,                
	    colorlinks=true,                  
	    linkcolor=purple,                 
	    citecolor=teal,                   
	    filecolor=black,                  
	    urlcolor=blue                     
	}

\usepackage{calc}
\usepackage{longtable}
\usepackage{empheq}     
\usepackage[font=small,labelfont=bf]{caption} 
\usepackage{subcaption} 
\usepackage{siunitx}    
\usepackage{mathtools}
\usepackage{stmaryrd}   
\usepackage{dsfont}
\usepackage{bm}         
\usepackage{booktabs}   

\definecolor{mygreen}{RGB}{28,172,0} 
\definecolor{mylilas}{RGB}{170,55,241}
\definecolor{mygray}{rgb}{0.4,0.4,0.4}
\definecolor{myorange}{rgb}{1.0,0.4,0}

\def\restriction#1#2{\mathchoice
              {\setbox1\hbox{${\displaystyle #1}_{\scriptstyle #2}$}
              \restrictionaux{#1}{#2}}
              {\setbox1\hbox{${\textstyle #1}_{\scriptstyle #2}$}
              \restrictionaux{#1}{#2}}
              {\setbox1\hbox{${\scriptstyle #1}_{\scriptscriptstyle #2}$}
              \restrictionaux{#1}{#2}}
              {\setbox1\hbox{${\scriptscriptstyle #1}_{\scriptscriptstyle #2}$}
              \restrictionaux{#1}{#2}}}
\def\restrictionaux#1#2{{#1\,\smash{\vrule height .8\ht1 depth .85\dp1}}_{\,#2}} 

\newcounter{rmq}[section]
\DeclareMathOperator{\curl}{\vec{curl}}
\DeclareMathOperator{\dive}{div}
\renewcommand{\vec}[1]{\mathbf{#1}}               
\newcommand{\tens}[1]{\underline{\underline{#1}}} 

\title{\textsf{Manufacturable blazed metasurface gratings\\designed by a 3D topology optimization model}}
\author[1,2]{Simon Ans}
\author[1]{Fr\'ed\'eric Zamkotsian}
\author[2]{Guillaume Dem{\'e}sy}
\date{}

\affil[1]{Aix Marseille Univ, CNRS, CNES, LAM, Marseille, France}
\affil[2]{Aix Marseille Univ, CNRS, Centrale Med, Institut Fresnel, Marseille, France}

\begin{document}
\maketitle
\begin{abstract} 
    We present an open-source \cite{diffGratTopOpt} Finite Element Method-based topology
    optimization framework for 3D blazed metasurfaces operating in reflection
    over the [400--${1500}$]\,nm visible and near-infrared range. The design
    region is described through a density-based interpolation and optimized
    using the adjoint method, enabling the treatment of several tens of
    thousands of degrees of freedom. A first approach directly applies topology
    optimization to the 3D Finite Element mesh (mesh-based), yielding a freeform
    structure that achieves an average diffraction efficiency of 62\% in
    the $-$1st order over two octaves under the targeted incidence. However,
    the freeform patterns obtained remain difficult to manufacture. We therefore
    introduce a pillar-based parameterization, embedding fabrication constraints
    within the optimization loop. The resulting binary metasurface, compatible
    with e-beam lithography and Reactive Ion Etching techniques, achieves an
    average efficiency of 57\% over the same spectral band in $p$-polarization,
    with low polarization dependence. This work demonstrates that large-scale 3D
    topology optimization can bridge the gap between broadband optical
    performance and realistic nanofabrication constraints for blazed
    metasurfaces.
\end{abstract}


\section{Introduction}

Spectrographs are widely used in physics, biology and medicine, for material
characterization purposes. Determining the composition of a material
using a non-invasive method is of paramount interest, as it allows for the
identification of molecules or atoms in living beings. This forms the basis of
disease diagnosis, for instance, and more generally, of all types of optical
quality control in environmental and industrial sectors. For Earth and Universe
Observation, spectroscopy allows us to identify atomic and molecular species, to
measure redshifts in deep-field observations, to characterize exoplanetary
atmospheres and to monitor ecosystems, mineral compositions, or pollution
phenomena.

Blazed gratings are fundamental optical components in spectrographs, especially
for studying low-flux radiation. The most common type, whose fabrication has
been industrialized for certain applications, is the sawtooth-profiled grating
(see Fig.~\ref{fig::blazing_effect}, in reflection). The grating period is
chosen above the considered wavelengths to deflect light on diffraction orders,
while its \textit{blazed} geometry favors a single order to increase the
signal-to-noise ratio (SNR). For visible (Vis) and Near Infrared (NIR) study,
the pattern characteristic size of a single period is therefore around several
micrometers. Although they demonstrate a high-level and robust blazing effect on
one octave --~from one wavelength to its double~-- the lack of potential
for further performance improvements has driven the community toward developing
subwavelength structures, commonly referred to as \textit{metasurfaces} in the
literature.

The first studies on blazed metasurfaces date back to the early 1990s and were
published in Ref.~\cite{Farn_graded_index}. The concept was later popularized by
P. Lalanne et al. through the design and fabrication of a blazed metasurface
operating in Vis and NIR, which they compared to a sawtooth blazed
grating~\cite{blazed_lalanne}. The promising results they obtained led the
authors to discuss the natural improvements of their work. On the one hand, they
suggested using computational electromagnetics to solve Maxwell’s equations for
a given structure. Two widely used numerical methods are the Rigorous Coupled
Wave Analysis (RCWA)~\cite{RCWA_binary_gratings} and the Finite Element Method
(FEM)~\cite{book_fresnel_c5}, both allowing the resolution of Maxwell's equations
in harmonic regime. The latter is chosen in this article because it 
relies on an unstructured mesh and thus offers more design flexibility than RCWA, 
in which geometries are restricted to vertical edges. 
On the other hand, the authors invited the reader to locally adjust 
the structure's geometry and improve its performance through mathematical and 
computational techniques.

The numerical resolution of PDEs with the FEM has been a major breakthrough in
Computational Physics and Engineering. Physical problems involving realistic
geometries can be solved numerically in many fields, such as fluid mechanics
(Navier-Stokes equation), solid mechanics (convection-diffusion equations,
dynamics equations), thermodynamics (heat equation), acoustics (wave equation)
and of course photonics (Maxwell’s equations). The idea of modifying the
discretized domain to optimize its performance followed naturally. By
"performance", we mean a given \textit{objective function} to be improved
(\textit{e.g.}, the resistance of a structure or efficiency of a device), which
is also called target or merit function. The concept of structural optimization
based on FEM appeared in the 1960s, introduced by L. Schmit’s group according to
Ref.~\cite{structopt_history}. Modifying the \textit{design region} and its
so-called \textit{discretization mesh} constitutes shape optimization, which
remains widely used today. The works of G.~Allaire~\cite{allaire_shapeopt} are
often cited as a reference in this domain.

The challenges associated with shape optimization, such as mesh
adaptation and sensitivity analysis --~which have been extensively
addressed~\cite{mesh_evolution,ShapeOpt_Geuzaine}~-- led to the emergence of
another optimization method in the late 1980s: Topology Optimization (TO),
introduced in Ref.~\cite{topopt_origins}. Instead of modifying shape boundaries,
TO finds the optimal design by directly changing the material distribution
within small elementary "boxes", called voxels (a blend of "volume" and
"pixel"). These voxels are defined once and for all at the beginning of the
computation, allowing the Finite Element (FE) mesh to remain fixed. The numerous
improvements of this method, mainly attributed to M.~P.~Bends{\o}e,
O.~Sigmund~\cite{topOpt_Sigmund} and their collaborators, have established TO as
a powerful and versatile design framework. At first developed for solid
mechanics, it has now been applied to any field governed by PDEs.  A
non-exhaustive list includes fluid mechanics and fluid-structure
interactions~\cite{sweden_fluid_opt,IFS_topOpt},
acoustics~\cite{mechanicsOpt_Jensen, acoustic_cool},
electromechanics~\cite{mixedOpt_Geuzaine, electromech_sigmund} and
photonics~\cite{topdesign_friis,adjointAnalysis,Brazil_opt}. The latter is the
domain of interest for this work.

More specifically, TO in nanophotonics has been largely developed thanks to the
heuristic homogenization of the density parameters, called the Solid Isotropic
Material with Penalization method (SIMP), introduced by Bends{\o}e and
Sigmund~\cite{SIMP_sigmund}. Noticeable examples are the open-source
optimization repository MetaNet~\cite{metanet} and Machine Learning (ML)-aided
optimization tools~\cite{InverseDesign,adibi_hilab}. To our knowledge, the
state-of-the-art performance of manufactured broadband blazed metasurface
gratings designed with TO is around 55\% of average efficiency for 5 wavelengths
between 900\,nm and ${1300}$\,nm, for both polarizations (different grating for
TE and TM)~\cite{Sell_Fan_multiwavelength}. Another design, solely based on the
phase shift induced by blazed gratings~\cite{nanoblazed_IOF}, has demonstrated
the same performance on the wavelength range [350--${1100}$]\,nm. There are some
recent works in the IR with very high broadband performances, such as in
Ref.~\cite{broadband_IR}. However the targeted wavelength range of
[6--12]\,$\mu$m in this study is actually an octave, with the same energy gap as
the [400--800]\,nm visible range. Broadband studies in the Vis/NIR region
remain scarce.

This is why we have recently developed an open-source, bespoke, FEM-based
large-scale topology optimization tool that addresses the issue of
nanostructured blazed gratings for the Vis/NIR range
[400--${1500}$]\,nm~\cite{ownArticle}, based on Python and the Gmsh/GetDP
software suite~\cite{gmsh,getdp}. It can find an optimal design on any
wavelength range, on any diffraction order, with any pair of materials under
conical incidence, \textit{i.e.} mono-periodic (or 1D) grating with a free
incidence. We previously demonstrated a silica design with an average
efficiency of 81\% on these two octaves, hence 56\% higher than the current
sawtooth-profiled gratings (52\%) in relative terms. Its manufacturability was
however compromised, since it contained levitating elements and highly
challenging elementary patterns. Dual material gratings ensuring mechanical
sustainability had been tested with the same solver~\cite{ownProc_SPIE}. The
performances were unfortunately degraded, while the fabrication remained highly
challenging.

In this article, we generalize our open-source design solver to the 3D case and
adapt it to manufacturability constraints established in
Ref.~\cite{ownProc_ICSO}. In particular, we present a manufacturable metasurface
made of elementary pillars that outperforms the current broadband proposals
on the two octaves [400--$\num{1500}$]\,nm. The theoretical general background
of bi-periodic (or 2D) gratings is introduced in
Section~\ref{sec::theoretical_background}. The TO problem and its resolution
\textit{via} the adjoint method are exposed in
Section~\ref{sec::opt_problem}. The two improvements developed on our
optimization tool are presented in Section~\ref{sec::two_ways}. A 3D mesh-based
solution is described in Section~\ref{sec::mesh_based_3D} and a pillar-based
solution in Section~\ref{sec::pillar_based_3D}. Conclusion and perspectives are
given in Section~\ref{sec::conclusion}. The code in the conical/3D mesh-based and
3D pillar-based frameworks is totally open source and available in the
\href{https://gitlab.onelab.info/doc/models/-/tree/master/DiffractionGratingsTopOpt}{Code
1}. The comparison between the conical and the 3D mesh-based solvers, completing
Section~\ref{sec::mesh_based_3D}, is detailed in \hyperlink{app::appendix}{Appendix}.

\section{Set up of the 3D electromagnetic problem}\label{sec::theoretical_background}

\subsection{Problem Description}

The device that deflects each wavelength inside a spectrograph is a
\textit{diffraction grating}. It is a periodic structure, exhibiting a density
of tens to several thousands of lines per millimeter for visible and NIR optics.
Depending on the instrument, diffraction can occur either in transmission or in
reflection.

\begin{figure}[ht!]
    \centering
    \includegraphics[width=1\textwidth]{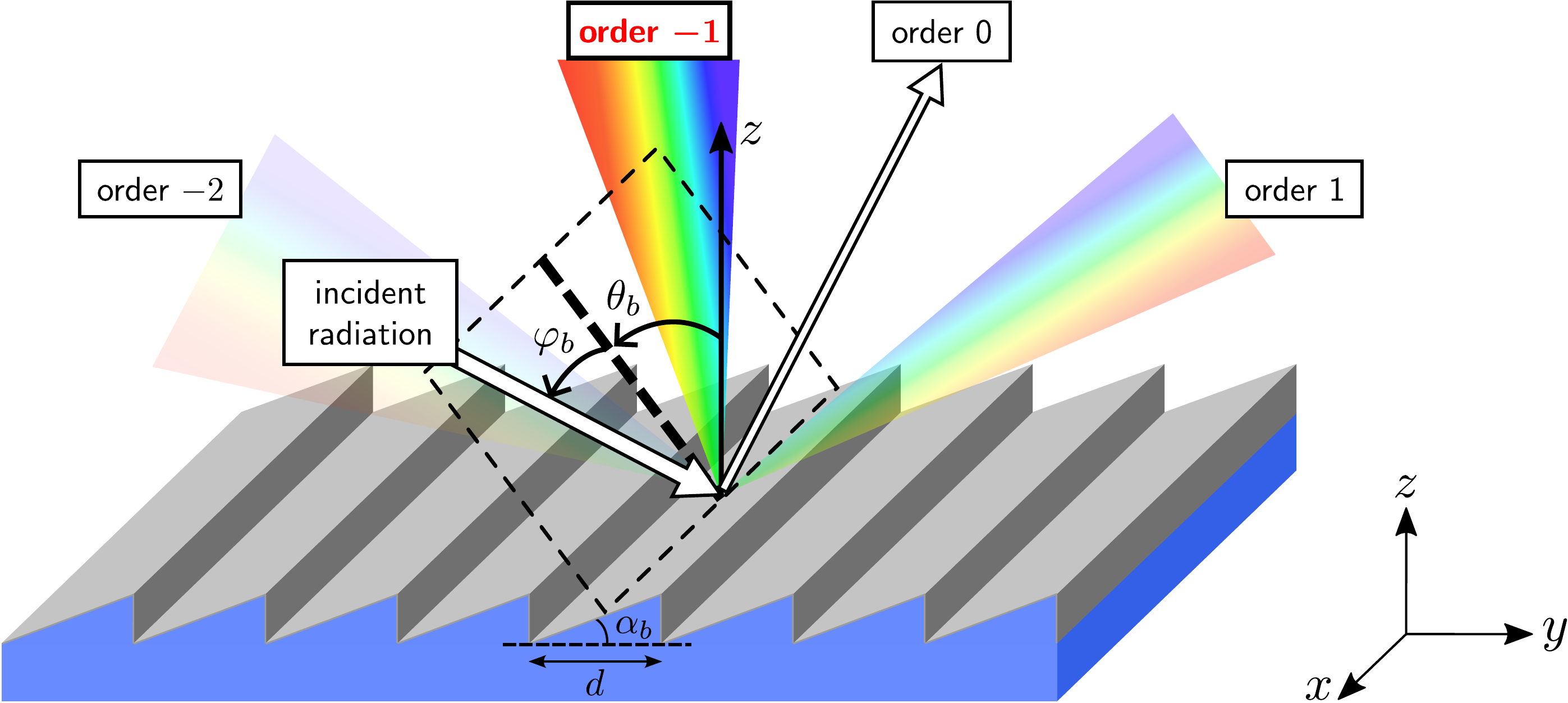}
    \caption{The blazing effect of a sawtooth grating working in reflection,
    with a white light incident radiation. The diffraction phenomenon is
    illustrated with 4 diffraction orders: $-2$, $-1$, 0 and 1. The grating is
    blazed, \textit{i.e.}, the majority of the diffracted light is
    directed into a specific order (here the $-1$st).}
    \label{fig::blazing_effect}
\end{figure}

Without specific optimization of the grating shape, the diffraction
efficiency is distributed among several propagative orders. However, to obtain
scientifically usable images --~that is, to reach a desired SNR~--
maximizing the flux into a single diffraction order is preferred. Gratings that
achieve this are known as \textit{blazed gratings} and the traditional sawtooth
profile is illustrated in Fig.~\ref{fig::blazing_effect}.

\begin{figure}[ht!]
    \centering
    \includegraphics[width=1.\textwidth]{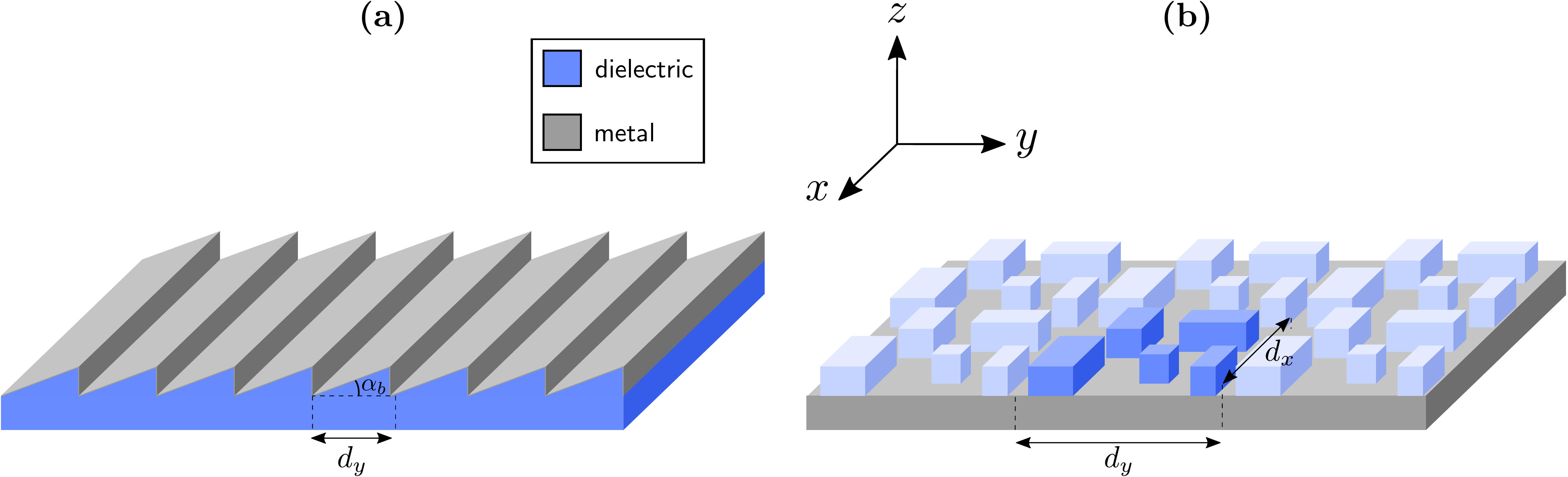}
    \caption{\textbf{(a)}~The reference sawtooth blazed grating. It is a
    mono-periodic (or 1D) grating with a period denoted $d_y$ and the so-called
    blaze angle $\alpha_b$. \textbf{(b)}~Pillar-shaped bi-periodic (or 2D)
    metasurface grating. One period of the pattern is highlighted with a darker
    color; its surface is $d_x \times d_y$.}
    \label{fig::triangle_metasurface}
\end{figure}

Our goal is to use the high versatility of the metasurface to optimize its 3D
geometry, as has already been done in 2D in Ref.~\cite{ownArticle}. On the one
hand, the sawtooth grating shown in
Fig.~\ref{fig::triangle_metasurface}a leaves little room for improvement.
Changing the combination of the \textit{blaze angle} $\alpha_b$ and the period
$d_y$ only changes the targeted wavelength range and diffraction order. On the
other hand, the bi-periodic or 2D metasurface configuration sketched in
Fig.~\ref{fig::triangle_metasurface}b (here with a \textit{binary grating}
composed of dielectric pillars) demonstrates a wide range of parameters subject
to optimization --~layer thickness, pillar dimensions, material. Since the
number of optimization Degrees of Freedom (opt. DoFs) is large, and to
maintain a constant mesh throughout the process, TO is chosen.

A more general definition of bi-periodic gratings is given in the next
subsection, enabling a freeform optimized pattern and preparing the
optimization process.

\subsection{Computational Domain in 3D}

A point in the domain is denoted $\vec{x} = (x,y,z) \in \mathbb{R}^3$. The
($Oz$) axis is normal to the design region. Therefore, the main period of the
grating is along the ($Oy$) axis and is thus denoted $d_y$; the grating lines
are along the ($Ox$) axis and the subwavelength period is thus denoted $d_x$. 

\begin{figure}[ht!]
    \centering
    \includegraphics[width=1.\textwidth]{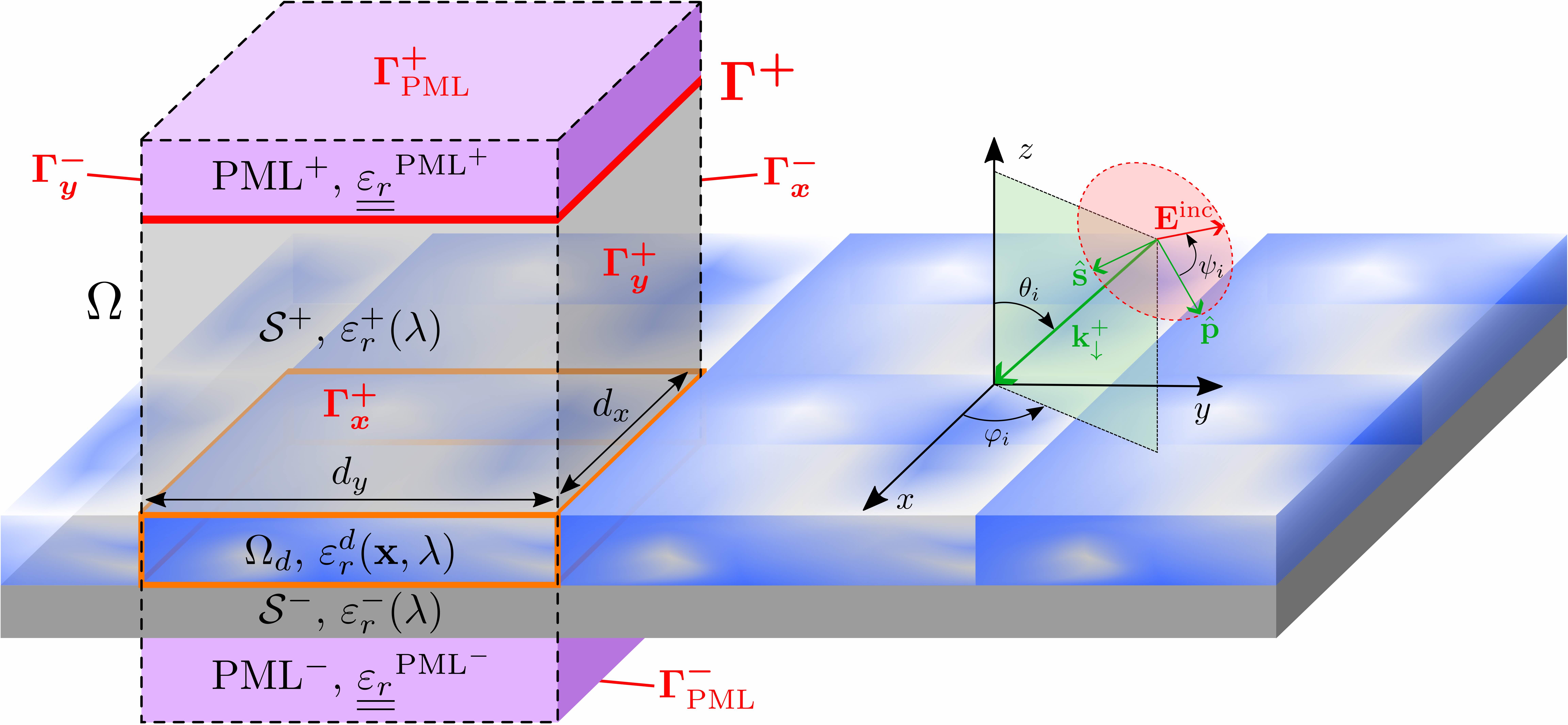}
    \caption{Numerical domain for a bi-periodic grating and definition of the
    incident electric field. In each domain, the permittivity tensor field is
    defined. The region of interest (where the metasurface lies) is the design
    region $\Omega_d$ (surrounded by the orange cuboid). The material where the
    incident radiation travels is the superstrate $\mathcal{S}^+$ (in
    transparent gray). The reflecting metallic layer is the substrate denoted
    $\mathcal{S}^-$ (in gray). The interface on top of $\mathcal{S^+}$, called
    $\Gamma^+$ (in red), is where the diffracted field is analyzed. The
    periodicity of the pattern allows the computational domain to be restricted to a
    single $d_x\times d_y$ period, with absorbing Perfectly Matched Layers up
    and down ($\text{PML}^{\pm}$, in violet). The incident electric field is
    perpendicular to the direction of propagation given by the vector
    $\vec{k}^+_{\downarrow}$. It is defined by the polar angle $\theta_i$, the
    azimuthal angle $\varphi_i$ and the polarization angle $\psi_i$.}
    \label{fig::numerical_domain}
\end{figure}

The structure and numerical domain parameters are illustrated in
Fig.~\ref{fig::numerical_domain}. One finds from top to bottom: the top
Perfectly Matched Layer (PML$^+$), absorbing the outgoing waves; the superstrate
$\mathcal{S}^+$; the design region $\Omega_d$; the substrate $\mathcal{S}^-$;
the bottom PML PML$^-$, which is not necessary when working in reflection with a
metallic substrate. All sub-domains are parallelepipeds. For reflected
(respectively (resp.) transmitted) diffracted fields, we make use of the top
interface $\Gamma^+$ highlighted in red between $\mathcal{S}^+$ and PML$^+$
(resp. $\Gamma^-$ between $\mathcal{S}^-$ and PML$^-$).

The grating is illuminated by an incident linearly polarized plane wave
$\vec{E}^{\text{inc}}$ with a free-space wavelength $\lambda$ (frequency $f$,
angular frequency $\omega$). The angles of incidence $\theta_i \in [0,\pi/2[\,$,
$\varphi_i \in \,]-\pi,\pi]$ define its wavevector $\vec{k}^+_{\downarrow}$ in
spherical coordinates. The explicit definition of $\vec{k}^+_{\downarrow}$ is
given later in the next subsection. The polarization basis vectors
$\hat{\vec{s}}$ and $\hat{\vec{p}}$ are defined such that $(\hat{\vec{p}},
\hat{\vec{s}}, \vec{k}^+_{\downarrow}/k^+)$ forms an orthonormal basis ($k^+
\coloneqq \lvert \vec{k}^+_{\downarrow} \rvert $) and $\hat{\vec{s}}$ is
parallel to the $(Oxy)$ plane. The polarization angle $\psi_i$, between
$\hat{\vec{p}}$ and $\vec{E}^{\text{inc}}$, allows one to define an
arbitrary linearly polarized plane wave.

Let $\rho: \Omega_d \ni \vec{x} \mapsto \rho(\vec{x}) \in [0,1]$ be a density
field in the design region and let $\vec{x}\in\Omega_d$ be a given point in this
same region. The relative permittivity at this position is defined using the
SIMP method, \textit{i.e.} the following bijection:
\begin{equation}\label{eq::density}
    \varepsilon_r^d(\lambda,\vec{x})
    = (\varepsilon_{r,2}(\lambda) - \varepsilon_{r,1}(\lambda))\rho(\vec{x}) + \varepsilon_{r,1}(\lambda)
\end{equation}
where $\varepsilon_{r,1}$ and $\varepsilon_{r,2}$ are resp. the minimal and
maximal relative permittivity in $\Omega_d$. Note that if one desires a complex
refractive index within the design region --~\textit{e.g.}, a metal or silicon
for UV/Vis light~-- a non-linear interpolation scheme demonstrates better
convergence capabilities~\cite{design_metal, ownArticle}.

In the next section, we briefly recall the 3D scattered field formulation
\cite{ownArticle} in the frame of finite elements.

\subsection{The Scattered Field Formulation in 3D}

The incident field is defined as $\vec{E}^{\text{inc}} = \vec{A}_e \exp(\iota
\vec{k}_{\downarrow}^+ \cdot \vec{x})$, $\forall \vec{x}\in \mathbb{R}^3$,
where~\cite{book_fresnel_c5}
\begin{equation}\label{eq::incident_wavevector}
    \vec{k}_{\downarrow}^+ =
    \begin{pmatrix}
        k_x \\ k_y \\ k_z^+
    \end{pmatrix} =
    k^+\begin{pmatrix}
        -\sin\theta_i\cos\varphi_i \\
        -\sin\theta_i\sin\varphi_i \\
        -\cos\theta_i
    \end{pmatrix},
\end{equation}
denoting $k^+ = k_0\sqrt{\varepsilon_r^+\mu_r^+}$ with $k_0 = 2\pi/\lambda$. The
vector amplitude writes
\begin{equation}\label{eq::amplitude}
    \vec{A}_e = A_e
            \begin{pmatrix}
                \cos\psi_i\cos\theta_i\cos\varphi_i - \sin\psi_i\sin\varphi_i \\
                \cos\psi_i\cos\theta_i\sin\varphi_i + \sin\psi_i\cos\varphi_i \\
                -\cos\psi_i\sin\theta_i
            \end{pmatrix}.
\end{equation}
We look for the scattered field $\vec{E}_*^d$, sum of all the outgoing fields of
the problem induced by the grating. As a consequence of the Floquet-Bloch
theorem, the response of a bi-periodic grating to a plane incident wave is a
bi-quasi-periodic wave, \textit{i.e.}, a field $\vec{U}$ such that
\begin{equation}\label{eq::biquasiperiodic}
    \forall\vec{x}\in \mathbb{R}^3, \quad \vec{U}(x,y,z) = \vec{U}^{\#}(x,y,z)e^{\iota k_x x}e^{\iota k_y y}
\end{equation}
where $\vec{U}^{\#}$ is a bi-periodic function with periods $d_x$ along $x$ and
$d_y$ along $y$. The Bloch boundary conditions $\vec{E}_*^d(\Gamma_{\{x,y\}}^+)
= e^{\iota k_{\{x,y\}} d_{\{x,y\}}} \vec{E}_*^d(\Gamma_{\{x,y\}}^-)$ are
therefore enforced.

In addition, if the PMLs have a sufficient thickness, the field can be safely
neglected on their far end (at the maximal |z|). By "sufficient thickness", we mean
a distance for which $\lvert \vec{E}_*^d \rvert \ll \lvert \vec{E}^{\text{inc}}
\rvert$ at the PML's ending. Typically, $2\lambda_{\min}$ is satisfactory
enough, with in our case $\lvert \vec{E}_*^d \rvert \simeq 10^{-5} \lvert
\vec{E}^{\text{inc}} \rvert$. A Dirichlet condition is therefore
imposed on the upper and lower boundaries to reduce the number of FEM unknowns:
\begin{equation}\label{eq::tangential_Dirichlet}
    \vec{E}_*^d \times \hat{\vec{z}} = 0 \text{ in } \Gamma_{\text{PML}}^{\pm}.
\end{equation}
Let $\boldsymbol{\mathcal{V}} \coloneqq \boldsymbol{\mathcal{H}}_0(\curl,
\Omega, k_x d_x, k_y d_y)$ be the Sobolev space with these
boundary conditions --~0 designating tangential Dirichlet and $k_x d_x$, $k_y
d_y$ the bi-quasi-periodicity. The scattered field formulation of the Helmholtz
propagation equation is:
\medbreak
\noindent Find $\vec{E}^d_* \in \boldsymbol{\mathcal{V}}$ such that for all
$\vec{x}\in\mathbb{R}^3$ with $\omega\in\mathbb{R}^+$,
\begin{equation}\label{eq::scattered_field_PDE}
    \curl \Big(\tens{\mu_r}^{-1} \curl\vec{E}_*^d \Big)
    - k_0^2 \tens{\varepsilon_r}\vec{E}_*^d
    = k_0^2 \Big( \tens{\varepsilon_r} - \tens{\varepsilon_r^a} \Big) \vec{E}_a,
\end{equation}
where $\vec{E}_a$ is the solution of the ancillary problem $\curl
\Big(\tens{\mu_r}^{-1} \curl\vec{E}_a \Big) - k_0^2
\tens{\varepsilon_r^a}\vec{E}_a = 0$ and $\varepsilon_r^a$ is the distribution
of permittivity without the groove region --~only the superstrate/substrate
interface. For more comprehensive information on this scattered field formulation
in presence of a substrate, one can refer to Ref.~\cite{book_fresnel_c5}.
The associated weak formulation is finally :
\medbreak
\noindent Find $\vec{E}^d_* \in \boldsymbol{\mathcal{V}}$ such that for all
$\vec{E}' \in \boldsymbol{\mathcal{V}}$,
\begin{equation}\label{eq::weak_formulation}
        \int_{\Omega}
        \Biggl[
        \tens{\mu_r}^{-1} \curl\vec{E}_*^d \cdot \curl\overline{\vec{E}'}
        - k_0^2 \left(\tens{\varepsilon_r}\vec{E}_*^d
                    - (\tens{\varepsilon_r^a} - \tens{\varepsilon_r})\vec{E}_a \right) \cdot \overline{\vec{E}'}
        \Biggr]
        \, \mathrm{d}\Omega
        = 0.
\end{equation}

\subsection{Diffraction efficiencies}

The components of the wave vectors for each diffraction order in reflection
$(n,m)\in \{(n,m) \in \mathbb{Z}^2 \, / \, k_{z,n,m}^+ \in \mathbb{R}\}$ (resp.
$k_{z,n,m}^-\in\mathbb{R}$ in transmission) are then:
\begin{equation}\label{eq::wave_vectors_diffraction}
    \left\{
    \begin{array}{llc}
        k_{x,n}     & = k_x + \dfrac{2\pi}{d_x}n \\
        k_{y,m}     & = k_y + \dfrac{2\pi}{d_y}m \\
        k_{z,n,m}^+ & = \sqrt{k_0^2\varepsilon_r^+ - k_{x,n}^2 - k_{y,m}^2} \\
        k_{z,n,m}^- & = -\sqrt{k_0^2\varepsilon_r^- - k_{x,n}^2 - k_{y,m}^2}
    \end{array}
    \right. .
\end{equation}
The so-called complex amplitudes in reflection $r_{n,m}^u$ and in transmission
$t_{n,m}^u$ ($u\in\{x,y\}$), coming from the Fourier decomposition, write
\begin{equation}\label{eq::complex_amplitudes}
    \left\{
    \begin{array}{llc}
        \displaystyle r_{n,m}^u = \dfrac{1}{d_x d_y}\int_{\Gamma^+} e^{-\iota(k_{x,n} x + k_{y,m} y)}\vec{E}^d(x,y,z_0)\cdot\hat{\vec{u}}\,\mathrm{d}x\mathrm{d}y & \mbox{for } z_0 > z_{\Omega_d} \\
        \displaystyle t_{n,m}^u = \dfrac{1}{d_x d_y}\int_{\Gamma^-} e^{-\iota(k_{x,n} x + k_{y,m} y)}\vec{E}^{\text{tot}}(x,y,z_0)\cdot\hat{\vec{u}}\,\mathrm{d}x\mathrm{d}y & \mbox{for } z_0 < 0
    \end{array}
    \right. .
\end{equation}
Finally the diffraction efficiencies are computed using the formula in
Ref.~\cite{theseZolla}, using only the continuous components of the field
--~\textit{i.e.}, tangential to the interfaces, hence along $x$ and $y$:
\begin{equation}\label{eq::efficiencies}
    \left\{
        \begin{array}{llc}
            R_{n,m} = \dfrac{1}{k_{z,n,m}^+ k_z^+ A_e^2}
            & \Bigl[(k_{x,n}^2 + (k_{z,n,m}^+)^2)\lvert r_{n,m}^x\rvert^2 \\
            & + (k_{y,m}^2 + (k_{z,n,m}^+)^2)\lvert r_{n,m}^y\rvert^2 \\
            & + 2k_{x,n}k_{y,m}\text{Re}(r_{n,m}^y\overline{r_{n,m}^x})\Bigr] \\
            
            T_{n,m} = \dfrac{1}{k_{z,n,m}^- k_z^+ A_e^2}
            & \Bigl[(k_{x,n}^2 + (k_{z,n,m}^-)^2)\lvert t_{n,m}^x\rvert^2 \\
            & + (k_{y,m}^2 + (k_{z,n,m}^-)^2)\lvert t_{n,m}^y\rvert^2 \\
            & + 2k_{x,n}k_{y,m}\text{Re}(t_{n,m}^y\overline{t_{n,m}^x})\Bigr]
        \end{array}
    \right. .
\end{equation}
\begin{proof} The diffraction efficiency in reflection $R_{n,m}$ (resp.
    in transmission $T_{n,m}$) for a given diffraction order $(n,m)$ is defined
    as the normalized norm of the vector $(r_{n,m}^u)^{\top}_{u\in\{x,y,z\}}$
    (resp. $(t_{n,m}^u)^{\top}_{u\in\{x,y,z\}}$):
    \begin{equation}\label{eq::efficiencies_Poynting}
        \left\{
            \begin{array}{llc}
                R_{n,m} = \dfrac{k_{z,n,m}^+}{k_z^+ \lvert A_e\rvert^2} \Bigl[\lvert r_{n,m}^x\rvert^2
                                                                            + \lvert r_{n,m}^y\rvert^2
                                                                            + \lvert r_{n,m}^z\rvert^2
                                                                        \Bigr] \\
                T_{n,m} = \dfrac{k_{z,n,m}^-}{k_z^+ \lvert A_e\rvert^2} \Bigl[\lvert t_{n,m}^x\rvert^2
                                                                            + \lvert t_{n,m}^y\rvert^2
                                                                            + \lvert t_{n,m}^z\rvert^2
                                                                        \Bigr]
            \end{array}
        \right. .
    \end{equation}

    Gauss's law $\varepsilon_0\dive\tens{\varepsilon_r}\vec{E} = \rho$ in a
    region where $\varepsilon_r$ is constant (the superstrate and the
    substrate) provides $\dive\vec{E} = 0$. Therefore, for every order
    $(n,m)$, the following identity is verified:
    \begin{equation}\label{eq::div_null_complex_amplitudes}
        k_{x,n} r_{n,m}^x + k_{y,m} r_{n,m}^y + k_{z,n,m}^{\pm} r_{n,m}^z = 0.
    \end{equation}
    Then
    \begin{equation}\label{eq::value_r_nmz}
        \begin{alignedat}{2}
            & r_{n,m}^z && = -\frac{1}{k_{z,n,m}^+} \left(k_{x,n} r_{n,m}^x + k_{y,m} r_{n,m}^y \right) \\
            \implies & \lvert r_{n,m}^z \rvert ^2 && = \frac{1}{(k_{z,n,m}^{\pm})^2} \left| k_{x,n} r_{n,m}^x + k_{y,m} r_{n,m}^y \right|^2  \\
            & && = \frac{1}{(k_{z,n,m}^{\pm})^2}
            \left[ k_{x,n}^2 \lvert r_{n,m}^x \rvert ^2 + k_{y,m}^2 \lvert r_{n,m}^y \rvert ^2 + 2 k_{x,n} k_{y,m} \text{Re} ( r_{n,m}^y \overline{r_{n,m}^x} ) \right].
        \end{alignedat}
    \end{equation}
    The injection of Eq.~\eqref{eq::value_r_nmz} in
    Eq.~\eqref{eq::efficiencies_Poynting} provides Eq.~\eqref{eq::efficiencies}.
\end{proof}
With this expression of the diffraction efficiencies, one
can shed the dependency on the normal component of the electric field (here the
$z$ axis) and simply compute its $x$ and $y$ components resp. as $\vec{E}\cdot
\hat{\vec{x}}$ and $\vec{E}\cdot \hat{\vec{y}}$.

Our goal is to optimize the reflection efficiency $R_{0,-1}$. Note that in
practice, the period $d_x$ is voluntarily taken subwavelength to prevent
diffraction on the orders $n\ne 0$. This second periodicity only allows for more
freedom in the optimal design, and we simply refer to $R_{0,-1}$ as "the $-1$st
diffraction order".

\section{Optimization Problem and Adjoint Method}\label{sec::opt_problem}

\subsection{Merit Function and its Jacobian}

Let $\mathfrak{T}$ be the triangulation obtained using the FEM. There is a
finite number $\mathcal{N}_T$ of tetrahedra in the design region $\Omega_d$. Let
then $\mathcal{T}_i \in \mathfrak{T}_{\lvert\Omega_d}$,
$i\in\{1,\dots,\mathcal{N}_T\}$ be one of these tetrahedra. The
relative permittivity is defined per mesh element, meaning that in
Eq.~\eqref{eq::density}, $\forall \vec{x}\in\mathcal{T}_i, \, \rho(\vec{x}) =
\rho_i$, with $\rho_i$ a constant real number between 0 and 1. The opt. DoFs are
the densities $\rho_i$ within the design region, and $\boldsymbol{\rho}
= (\rho_1,\dots,\rho_{\mathcal{N}_T})$ is called the \textit{design variable}.
Note that within this framework, the mesh is fixed during the whole
optimization process, which is a considerable advantage of the adjoint method.

A connectedness filter $\boldsymbol{\rho}_f$~\cite{TD_3D_TopOpt} and a
binarization filter $\boldsymbol{\hat{\rho}}$~\cite{ADCode} are applied
(see Appendix~\ref{app::constraints} for a comprehensive description of the filters).
Combined with the SIMP-defined distribution of material, the
binarization filter enables a complete description of the evolving boundaries of
the design during the optimization process. The filtered design variable writes
then
\begin{equation}\label{eq::comp_epsilon}
    \hat{\boldsymbol{\rho}}_f: [0,1]^{\mathcal{N}_T} \ni \boldsymbol{\rho} \mapsto (\hat{\boldsymbol{\rho}} \circ \boldsymbol{\rho}_f)(\boldsymbol{\rho}) \in [0,1]^{\mathcal{N}_T}.
\end{equation}
Let $(n,m)$ be the targeted diffraction order to maximize. We denote by
$\mathcal{R}_{n,m}$ the reflection efficiency on the order $(n,m)$ of the
structure obtained using the filtered design variable: $\mathcal{R}_{n,m}:
[0,1]^{\mathcal{N}_T} \ni \boldsymbol{\rho} \mapsto (R_{n,m} \circ
\hat{\boldsymbol{\rho}} \circ \boldsymbol{\rho}_f)(\boldsymbol{\rho}) \in
[0,1]$. The optimization problem is then to minimize the merit function
$\mathcal{F}_{n,m} := 1 - \mathcal{R}_{n,m}$. The optimization problem writes:
\begin{equation}\label{eq::opt_prob}
    \begin{alignedat}{2}
        \underset{\boldsymbol{\rho}}{\min} &                && \mathcal{F}_{n,m}(\boldsymbol{\rho}) \\
        \text{such that}                   & \text{ \ \ \ } && 
        \left\{ \begin{array}{llc}
                    \mathcal{L}(\boldsymbol{\rho}, \vec{E}') = 0 & \mbox{$\forall \vec{E}' \in \boldsymbol{\mathcal{V}}$} \\
                    0 \leq \rho_i \leq 1 & \mbox{$\forall i\in \{ 1,\dots,{\mathcal{N}_T} \}$}
                \end{array}
        \right. ,
    \end{alignedat}
\end{equation}
where $\mathcal{L}$ designates the weak formulation in
Eq.~\eqref{eq::weak_formulation}, that is:
\begin{equation}\label{eq::constr_L}
    \begin{alignedat}{1}
        \mathcal{L} : \mathbb{R}^{\mathcal{N}_T} \times \boldsymbol{\mathcal{V}} \ni (\boldsymbol{\rho},\vec{E}') \mapsto
        &\int_{\Omega} \bigg[\tens{\mu_r}^{-1} \curl\vec{E}_*^d(\boldsymbol{\rho}) \cdot \curl\overline{\vec{E}'} \\
                    &- k_0^2\Big(\tens{\varepsilon_r}(\boldsymbol{\rho})\vec{E}_*^d(\boldsymbol{\rho}) + (\tens{\varepsilon_r}(\boldsymbol{\rho}) - \tens{\varepsilon_{r,a}})\vec{E}_a\Big) \cdot \overline{\vec{E}'}
                    \bigg] \mathrm{d}\Omega \in \mathbb{R}.
    \end{alignedat}
\end{equation}
Using the chain rule, the Jacobian of the merit function of
Problem~\eqref{eq::opt_prob} is:
\begin{equation}\label{eq::target_derivative_general}
    \frac{\partial\mathcal{F}_{n,m}}{\partial\boldsymbol{\rho}}(\boldsymbol{\rho})
    = \frac{\partial\boldsymbol{\rho}_f}{\partial\boldsymbol{\rho}}(\boldsymbol{\rho})
    \frac{\partial\hat{\boldsymbol{\rho}}_f}{\partial\boldsymbol{\rho}_f}(\boldsymbol{\rho}_f)
    \frac{\partial F_{n,m}}{\partial\hat{\boldsymbol{\rho}}_f}(\hat{\boldsymbol{\rho}}_f),
\end{equation}
where $F_{n,m} = 1 - R_{n,m}$. The differentiation of the first line in
Eq.~\eqref{eq::efficiencies} leads to, for all $i\in\{1,\dots,\mathcal{N}_T\}$:
\begin{equation}\label{eq::derivative_Rnm}
    \begin{alignedat}{1}
        \frac{\partial F_{n,m}}{\partial \hat{\rho}_{f,i}}
        = - \frac{\partial R_{n,m}}{\partial \hat{\rho}_{f,i}} = -\frac{2}{k_{z,n,m}^+k_z^+ A_e^2}\text{Re}
            \Biggl[&(k_{x,n}^2 + (k_{z,n,m}^+)^2)\frac{\partial r_{n,m}^x}{\partial \hat{\rho}_{f,i}}\overline{r_{n,m}^x} \\
                 + \, &(k_{y,m}^2 + (k_{z,n,m}^+)^2)\frac{\partial r_{n,m}^y}{\partial \hat{\rho}_{f,i}}\overline{r_{n,m}^y} \\
                 + \, &k_{x,n}k_{y,m}\Bigl(\frac{\partial r_{n,m}^y}{\partial \hat{\rho}_{f,i}}\overline{r_{n,m}^x}
                 + r_{n,m}^y\frac{\partial \overline{r_{n,m}^x}}{\partial \hat{\rho}_{f,i}}\Bigr)
            \Biggr].
    \end{alignedat}
\end{equation}

\subsection{The adjoint problem to access the derivatives of the complex amplitudes}

The derivatives of $r_{n,m}^u$ with respect to the $\hat{\rho}_{f,i}$, $i\in \{
1,\dots,{\mathcal{N}_T} \}$, around the equilibrium point
$\hat{\boldsymbol{\rho}}_{f,*}$ are given by
\begin{equation}\label{eq::adjoint_derivative_rn}
    \frac{\partial r_{n,m}^u}{\partial \hat{\rho}_{f,i}} (\hat{\boldsymbol{\rho}}_{f,*})
    = k_0^2 \int_{\mathcal{T}_i} (\varepsilon_{r,2} - \varepsilon_{r,1}) \vec{E}^{\text{tot}} (\boldsymbol{\rho}_*) \cdot \boldsymbol{\lambda}_{*,n,m}^u \, \mathrm{d}\Omega,
\end{equation}
where $\mathcal{T}_i \in \mathfrak{T}_{\lvert\Omega_d}$ is the tetrahedron of
the mesh element in the design region with the density $\hat{\rho}_{f,i}$, and
$\boldsymbol{\lambda}_*^u$ is the unique solution of the adjoint problem, noting
$\boldsymbol{\mathcal{V}}^{\text{adj}} = \boldsymbol{\mathcal{H}}_0(\curl,
\Omega, -k_x x, -k_y y)$ (note that the Bloch boundary conditions have
been inverted, suggesting a backward propagative plane wave):
\medbreak
\noindent Find
$\boldsymbol{\lambda}_{*,n,m}^u\in\boldsymbol{\mathcal{V}}^{\text{adj}}$ such
that for all $\boldsymbol{\lambda}' \in \boldsymbol{\mathcal{V}}^{\text{adj}}$,
\begin{equation}\label{eq::adjoint_problem}
    \int_{\Omega}
    \Biggl[
    \tens{\mu_r}^{-1} \curl\boldsymbol{\lambda}_{*,n,m}^u \cdot \curl\overline{\boldsymbol{\lambda}'}
    - k_0^2 \tens{\varepsilon_r}\boldsymbol{\lambda}_{*,n,m}^u \cdot \overline{\boldsymbol{\lambda}'}
    \Biggr]\,\mathrm{d}\Omega
    = \frac{1}{d_x d_y} \int_{\Gamma^+} e^{-\iota k_{x,n} x} e^{-\iota k_{y,m} y} \hat{\vec{u}} \cdot\overline{\boldsymbol{\lambda}'} \,\mathrm{d}\Gamma .
\end{equation}
Classically, this expression describes the adjoint field as a plane
wave coming from the targeted order $(n,m)$ and propagating toward the design
region, interfering with the forward field (as illustrated for instance in
Ref.~\cite{Sell_Fan_metanet}). This result enables the computation of
the merit function's Jacobian. A demonstration of this property can be found in
Ref.~\cite{ownArticle}. 

Finally, from the numerical point of view, the unknown electric field and adjoint variable 
are discretized by edge elements (full second order Webb elements \cite{webb1993hierarchal} 
are used throughout the paper, amounting to 20 FEM degrees of freedom per tetrahedron. The density field 
$\boldsymbol{\rho}$ is discretized using piecewise-constant $P_0$ elements. 
This discrete scheme allows for the expected tangential continuity of the electric field components 
across interfaces and elements, while the normal component is free to be discontinuous 
as the permittivity varies from one mesh tetrahedron to another.

\section{Mesh-based and pillar-based densities}\label{sec::two_ways}

\begin{figure}[ht!]
    \centering
    \includegraphics[width=.95\textwidth]{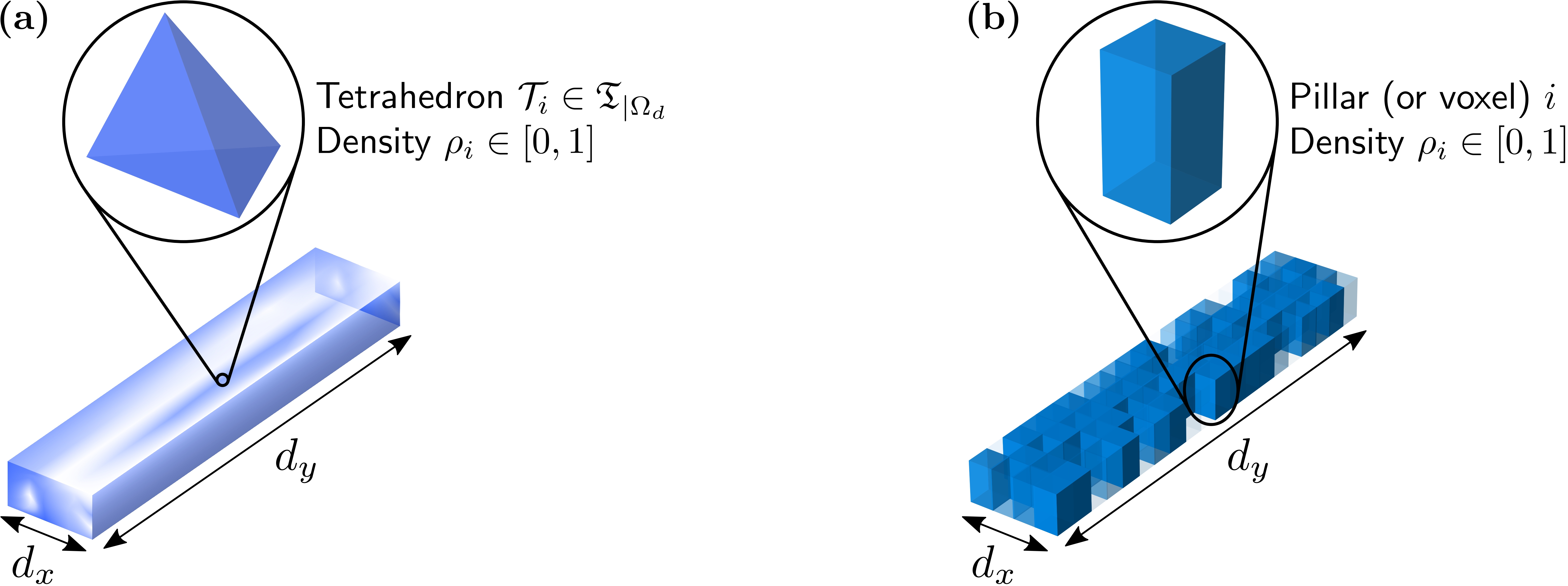}
    \caption{\textbf{(a)}~Mesh-based density distribution. The density $\rho_i$
    is optimized in every tetrahedron $\mathcal{T}_i \in
    \restriction{\mathfrak{T}}{\Omega_d}$. It can be seen as a freeform
    optimization (a tetrahedron is very small compared to the pattern size).
    \textbf{(b)}~Pillar-based density distribution. Same principle but the
    density is defined by cuboids with manufacturable sizes.}
    \label{fig::mesh_vs_pillar_based}
\end{figure}

For our study, the period $d_x$ is considered sub-wavelength to avoid the
diffraction along the grating lines (the orders corresponding to $n$ in
$R_{n,m}$). A period of the grating looks therefore like in
Fig.~\ref{fig::mesh_vs_pillar_based}, with $d_y \sim 10d_x$.

We present hereby two ways to define the density distribution using the SIMP.
The first restricts the support of each $\rho_i$ on a tetrahedron of the FE
mesh, as illustrated in Fig.~\ref{fig::mesh_vs_pillar_based}a. The second
defines $\rho_i$ on a manufacturable cuboid, like in
Fig.~\ref{fig::mesh_vs_pillar_based}b. By "manufacturable", we mean that the
dimensions of the elementary elements have been shown to be achievable with nanofabrication techniques
such as electron-beam (or more commonly e-beam)
lithography for the patterning and Reactive Ion Etching
(RIE)~\cite{explanation_RIE}. In this case, the geometry is meshed
independently of the definition of $\boldsymbol{\rho}$, and the meshing is only
used for the FEM forward and adjoint resolutions.

In both resolutions, we want to optimize a grating with $d_y =
3300$\,nm to obtain a blazing effect on the order $(n,m) = (0,-1)$ for the range
of wavelengths [400--$\num{1500}$]\,nm. The incident angles are $\theta_i =
24.5^{\circ}$, $\varphi_i = -11^{\circ}$ with a $p$-polarization $\psi_i =
0^{\circ}$. The angles $\theta_i$ and $\varphi_i$ are chosen to match the
specifications of an existing spectrometer~\cite{BATMAN_instrument}. 
These angles correspond to a physically consistent configuration expected to yield good performance.

In addition, the initial pattern of the optimization is always taken
as a linearly decreasing distribution of relative permittivity along the $y$
axis, independent from $x$ and $z$. For all $(x,y,z) \in \Omega_d$, we define
the initial density variable as $\rho_i = -(y-d_y) / d_y$.

\section{Mesh-Based Topology Optimization results}\label{sec::mesh_based_3D}

\begin{figure}[htb]
    \centering
    \includegraphics[width=1\textwidth]{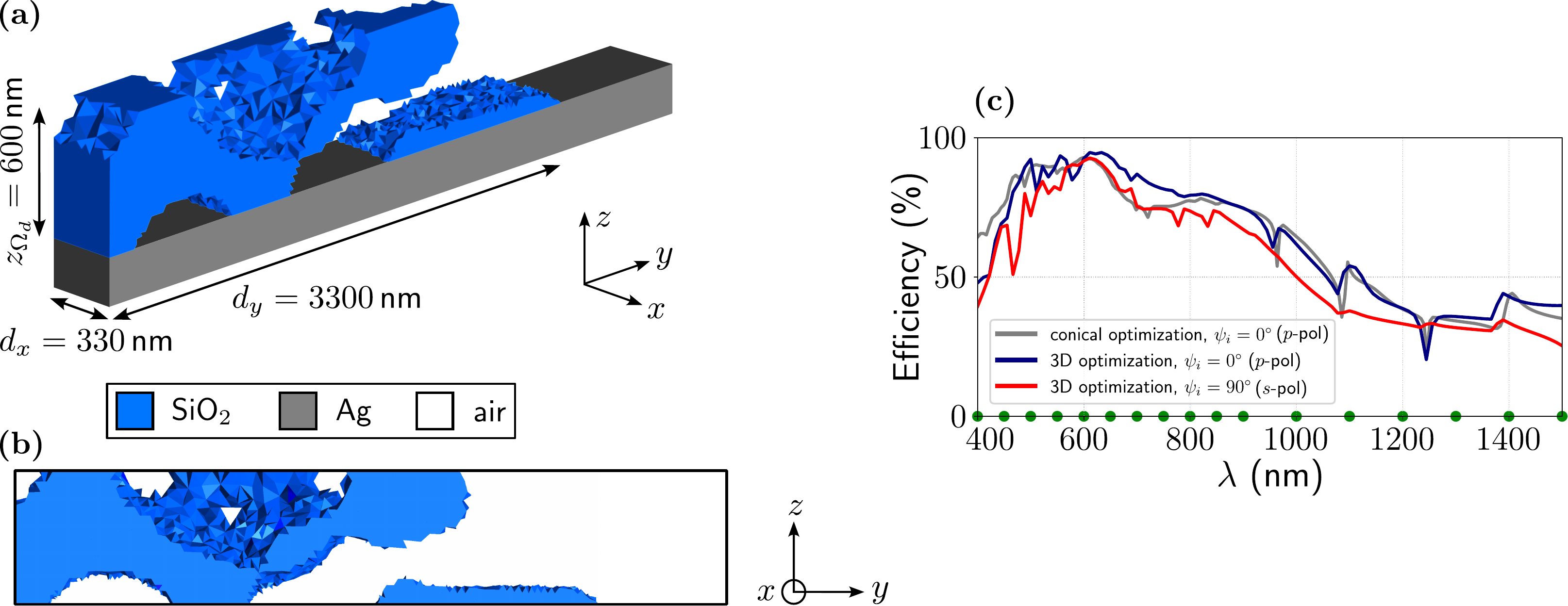}
    \caption{\textbf{(a)}~3D view of the optimal 3D pattern on a single period
    for a 600\,nm-thick design region with SiO$_2$ as chosen dielectric.
    \textbf{(b)}~Same pattern viewed on the plane $x=d_x$ (similar to the point
    of view of 2D gratings). \textbf{(c)}~Spectral response of the grating on
    the wavelength range of interest (blue line) as compared to the response for
    $s$-polarization ($\psi_i = 90^{\circ}$, red line) and the response for the
    targeted incidence of the conical optimal design (gray line, see
    Appendix~\ref{app::comparison}).}
    \label{fig::optimized_3D}
\end{figure}

The optimization on the 3D pattern is performed in a 600\,nm-thick design region
with $d_y=3300$\,nm and $d_x=330$\,nm, using a set of 17 wavelengths between
$\lambda=400$\,nm and $\lambda=1500$\,nm, with higher density in the visible
spectrum (see the targeted wavelengths represented by the green dots in
Fig.~\ref{fig::optimized_3D}c). The material is SiO$_2$. The $\num{47791}$ optimization
DoFs problem is launched on 17 threads for the wavelengths and 2 threads per
wavelength for the FE matrix inversion, on a workstation equipped with a 64-core
CPU and 4\,TB of RAM. The optimization process took 10 days for 882 iterations,
with RAM usage up to 1.2\,TB. Note that the precise number of optimization 
DoFs is directly determined by the unstructured FE meshing
algorithm, hence the seemingly arbitrary number of
tetrahedra obtained. In this case, with a mesh size of 
$r_T = \lambda_{\min}/(10\sqrt{\varepsilon_{r,2}(\lambda_{\min})})$ 
(\textit{i.e.} the minimal apparent wavelength in the design region divided by 10)
prescribed in the design region, the resulting number of tetrahedra 
(and of optimization DoFs) in the design region is $\num{47791}$.

The optimal pattern is shown with two points of view in
Figs.~\ref{fig::optimized_3D}a--b. The first view separates the three Cartesian
axes to provide a clear view of the structure's geometry and illustrates the
dimensions of the pattern's period. The second view reproduces the point of view
for 2D cases (on the $x=d_x$ plane) to highlight the striking resemblance
between this 3D profile and that of the optimal design in the conical case for
the same specifications (see Appendix~\ref{app::comparison}). Although this 3D
structure exhibits a variation of shape along the $x$ axis, the similarity is
confirmed with the spectral response of both patterns in
Fig.~\ref{fig::optimized_3D}c. The blue and gray lines correspond resp. to the
spectral response of the 3D and the conical grating. The gap between them
remains very narrow across the whole spectral range, and the average
efficiency is actually the same for both polarizations, at 62\%. This remains
higher than the state-of-the-art performances in this wavelength range found in
the literature. When using TO, they indeed demonstrate an average efficiency
around 55\% for the range [900--$\num{1300}$]\,nm, and two distinct designs are
obtained for $s$- and $p$-polarization~\cite{Sell_Fan_multiwavelength}. Without
using TO, the same 55\% average efficiency is obtained on the wavelength range
[350--$\num{1100}$]\,nm~\cite{nanoblazed_IOF}.

The benefit of our 3D design is that the polarization dependence is
significantly diminished with respect to the conical case (see
Appendix~\ref{app::comparison} for comparison). Now, the absolute average difference
between both polarizations on the 2 octaves is only 7\% --~the strong blazing
effect is more polarization-independent. This result is somewhat
surprising, as one might expect a superior local optimum given the additional
degrees of freedom enabled by the third dimension. Another improvement in this
optimal pattern is its better mechanical sustainability. Even if the grating is
not manufacturable with such dimensions, there is at least no levitating
isolated element.

The 3D mesh-based, topology-optimized grating is not the expected ultimate
design for high-efficiency blazing. Perhaps the roughness imposed by the mesh
tetrahedra --~more important in 3D as these elements are bigger than in the
conical case to limit the computational burden~-- is one of the causes of this lack of
improvement envisioned from the addition of a new dimension.

Topology optimization on bi-periodic structures is absolutely not abandoned,
however. If we can see the optimal pattern in the conical case as a maximal
threshold for mesh-based gratings, other discretization schemes of the design
variable $\boldsymbol{\rho}$ could bypass the strong roughness observed here and
maybe provide equivalent, or even better performance. To settle a framework for
these other distributions, the manufacturability of the grating becomes of
primary interest.

\section{Pillar-Based Topology Optimization results}\label{sec::pillar_based_3D}

\subsection{Choice of parameters for the pillars}

Commonly, topology optimization is seen as a method yielding a pattern with a
shape left free inside the design region. It is the case for the mesh-based
distribution of relative permittivity (Fig.~\ref{fig::mesh_vs_pillar_based}a)
since the mesh elements are small with respect to the filter radius $r_f =
300$\,nm. However, the method itself can be applied to discrete density
distributions other than the FE mesh elements. As motivated by the metasurface
fabrication, we now choose to rely on manufacturable cuboids as illustrated in
Fig.~\ref{fig::mesh_vs_pillar_based}b. Note that these cuboids are larger than
the typical mesh size, yet they remain smaller than the wavelength, preventing
sharp resonances inside a single voxel. In other words, increasing
the voxel size too much could lead to a poorer convergence of the adjoint
method, because the adjoint problem would not provide a correct
approximation of the merit function's derivatives.

From a manufacturing point of 
view, the pillar-based design relies on well-controlled fabrication process using
weakly resonant dielectric materials (silica, silicon nitride) rather than
structured metals. Many high-efficiency metasurfaces exploit sharp plasmonic
or high-index Mie resonances, which are extremely sensitive to nanometric
geometrical deviations and therefore difficult to fabricate reproducibly.
Working with low-index, weakly dispersive dielectrics yields a much smoother
dependence of the optical response on the geometry, which relaxes the
fabrication tolerances\cite{ans2025phd}.

\begin{figure}[h!tb]
    \centering
    \includegraphics[width=1\textwidth]{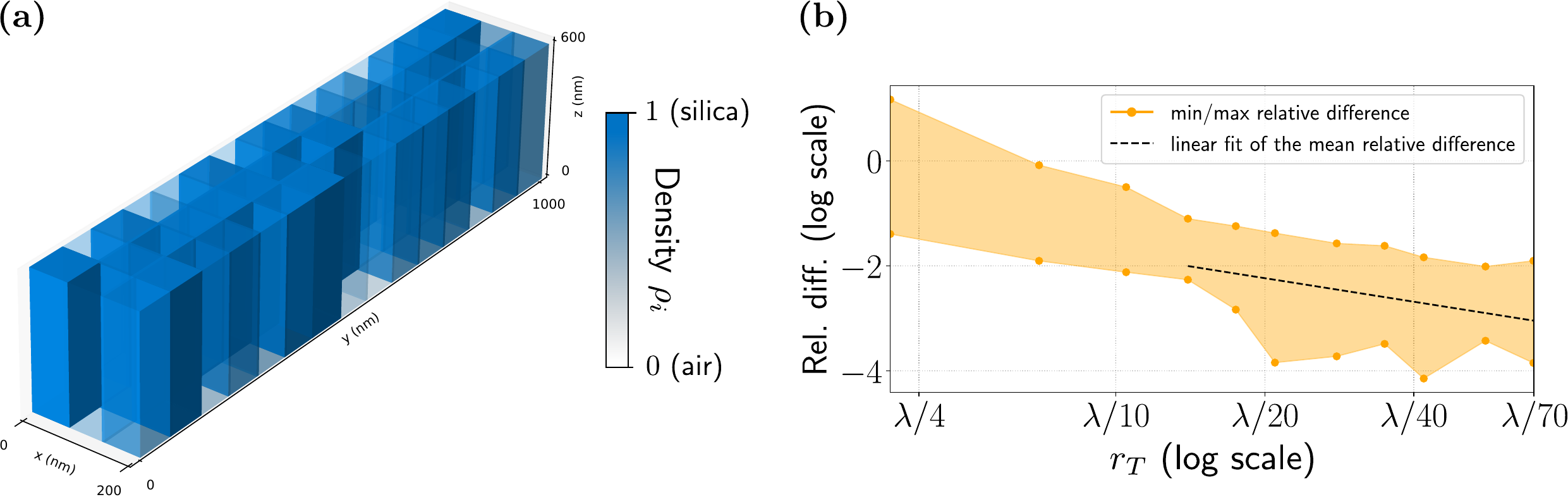}
    \caption{Convergence test of the adjoint method on the pillar-based grating,
    for random permittivities distributed on 45 squared pillars, on a $200\times
    {1000}\times 600$\,nm period, for the wavelength $\lambda = 700$\,nm and
    incident angles $\theta_i=24.5^{\circ}$, $\varphi_i=-11^{\circ}$ and
    $\psi_i=0^{\circ}$. \textbf{(a)}~Pillar-based random distribution obtained
    and set for the convergence test, with relative permittivities between 0
    (air) and 1 (silica) \textbf{(b)}~Minimum and maximum values (orange solid
    lines) taken by the relative differences between the adjoint-based and the
    finite-difference-based computations of the Jacobian of the merit function,
    delimiting the orange-colored region where all the relative values lie.
    Their mean value is linearly fitted (dashed black line) starting from $r_T =
    \lambda/14$.}
    \label{fig::convergence_pillar_based}
\end{figure}

A convergence test is thus provided, based on a numerical domain of
$200\times{1000}\times 600$\,nm. While this optimization method still relies on
a 3D model, the number of opt. DoFs is significantly lower. Therefore, comparing
the adjoint method to a simple finite difference performed on each pillar
without excessive processing time is possible.

Let us then define a random permittivity function, constant per pillar in a
$3\times 15$ pillar configuration and incident angles of
$\theta_i=24.5^{\circ}$, $\varphi_i=-11^{\circ}$ and $\psi_i=0^{\circ}$
($p$-polarization). This random permittivity function is kept the same for every
mesh size and is shown in Fig.~\ref{fig::convergence_pillar_based}a. We
represent in Fig.~\ref{fig::convergence_pillar_based}b the minimum and maximum
values (orange solid lines) of the relative difference between the adjoint-based
and the finite-difference-based computations of the Jacobian of the merit
function, in log$_{10}$ scales. The mean values of the relative differences on
the 45 pillars are linearly fitted starting from $r_T = \lambda/14$. It leads to
a gradient of $-1.5$, which highlights an average gain of 1.5 significant digits
on the accuracy of the adjoint method when the tetrahedron size is divided by
10.

In conclusion, the pillar-based optimization can still rely on the adjoint
method. The tetrahedron size is taken as $r_T = \lambda_{\min}/14$ to reduce the
computational burden while maintaining satisfactory precision. All the
manufacturability constraints introduced in this chapter are gathered in the
next subsection to provide a pattern that can be etched using e-beam lithography
and RIE.

\subsection{Manufacturable Metasurface Designed With Topology Optimization}

\begin{figure}[b!]
    \centering
    \includegraphics[width=.45\textwidth]{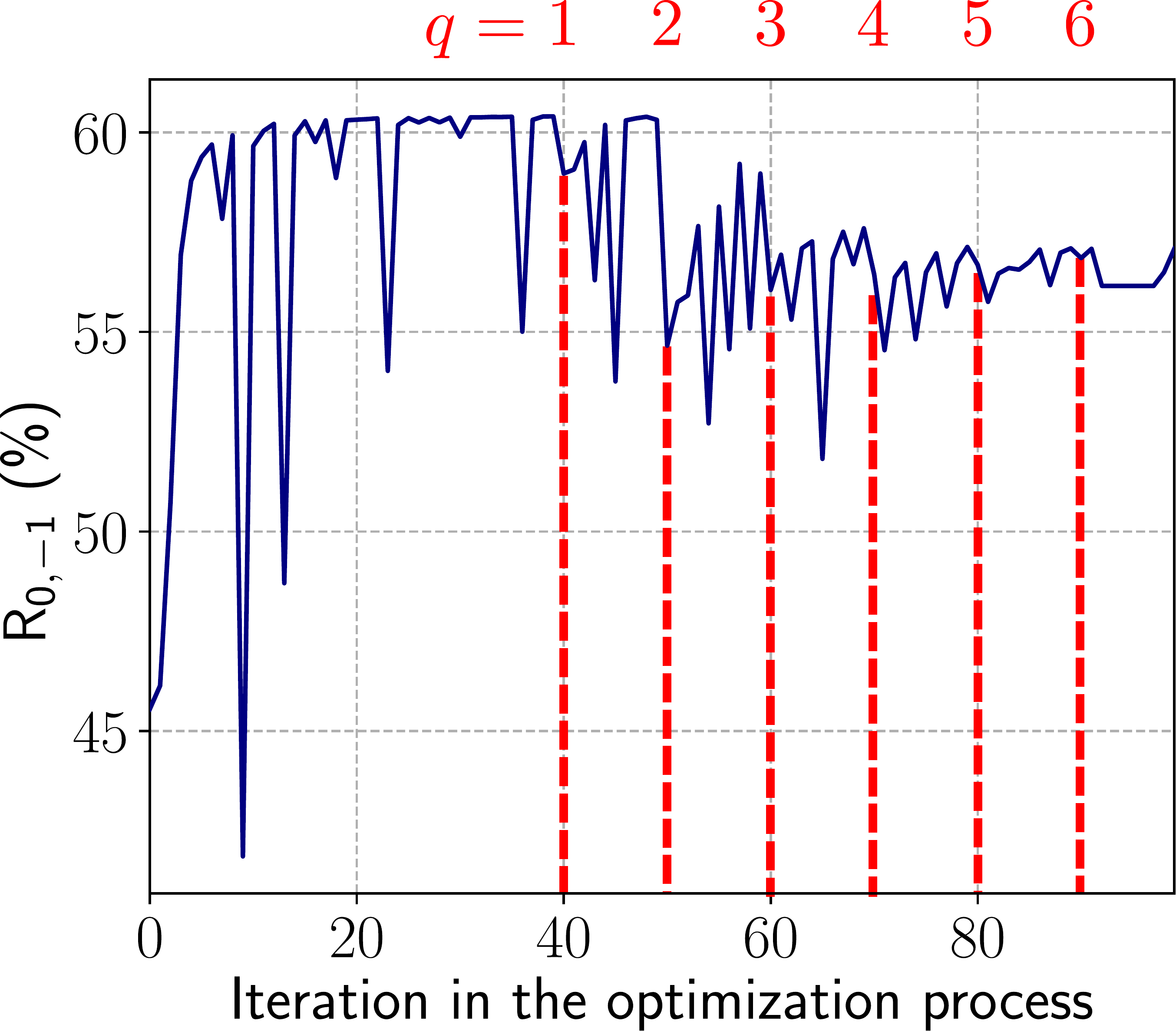}
    \caption{Convergence during the optimization process for a 200\,nm-thick,
    $4\times 40$ pillar pattern made of Si$_3$N$_4$ on a 50\,nm-thick layer of
    alumina. The binarization steps, identified by the binarization factor $q$
    (in red), are highlighted by the dashed red lines.}
    \label{fig::convergence_pillar_opt}
\end{figure}

In standard manufacturing processes including e-beam lithography and RIE, the
deposition of an etch stop layer protecting the silver substrate must be taken
into account. A 50\,nm-thick of alumina Al$_2$O$_3$ is therefore added in the
numerical domain and for the optimization process. The period of $330\times
{3300}$\,nm is divided into $4\times 40$ squared pillars, which leads to
lateral dimensions of 82.5\,nm. The dielectric is Si$_3$N$_4$ with a thickness
of $z_{\Omega_d}=200$\,nm, yielding an aspect ratio of 2.7 which is achievable
(see for instance the silica pattern in Ref.~\cite{nanoblazed_IOF} with an
aspect ratio of 3 and 600\,nm-thickness and the experimental realization in Ref.~\cite{ans2025phd}). 
The set of targeted wavelengths is
$\Lambda = \{400,500,600,\dots,{1500}\}$\,nm. The incident angles are
$\theta_i=24.5^{\circ}$, $\varphi_i=-11^{\circ}$ and $p$-polarization
$\psi_i=0^{\circ}$.

\begin{figure}[b!]
    \centering
    \includegraphics[width=1\textwidth]{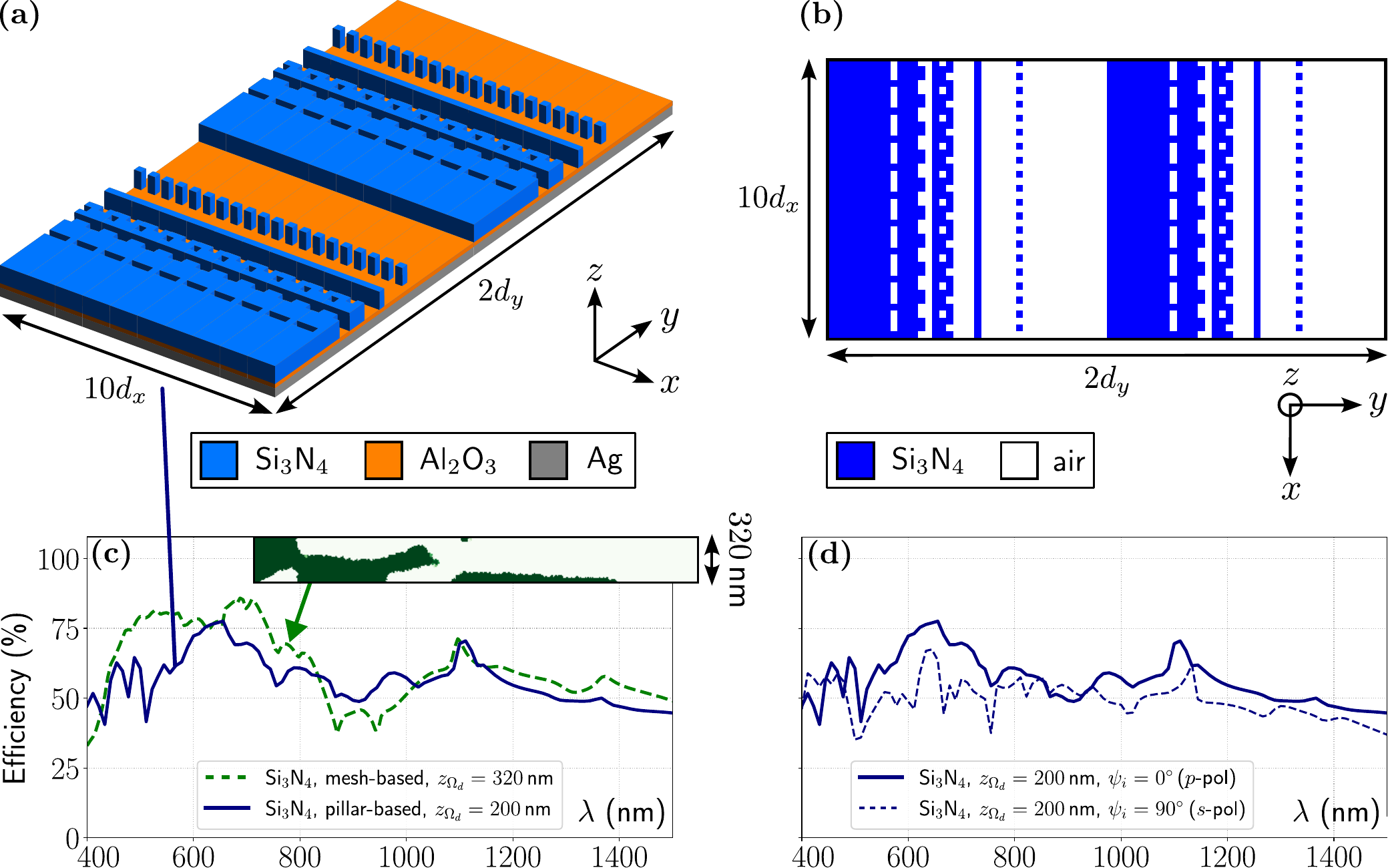}
    \caption{\textbf{(a)}~3D view of 10$\times$2 periods of the 200\,nm-thick
    Si$_3$N$_4$ optimal pattern. \textbf{(b)}~View from above of the same cell.
    Only Si$_3$N$_4$ (in blue) is represented. \textbf{(c)}~Spectral response in
    the $-1$st diffraction order of this pattern (solid blue line) as compared to
    that of the mesh-based 320\,nm-thick pattern under \textit{conical}
    incidence (pattern in green, spectral response in dashed green line,
    see Ref.~\cite{ownArticle} for the conical framework).
    \textbf{(d)}~Polarization dependence of the optimal pillar pattern for the
    $p$-polarization $\psi_i=0^{\circ}$ (solid line) and $s$-polarization
    $\psi_i=90^{\circ}$ (dashed line). 
    }
    \label{fig::pillar_performance}
\end{figure}

Reducing the number of opt. DoFs to only 160 is of great interest in 3D. The
convergence is much faster and one can set a lower maximum number of iterations.
Therefore, this optimization only takes 14\,h 50\,min to run 100
iterations, parallelized on wavelengths with 4 threads per wavelength (therefore
48 threads in total). The iteration history is shown in
Fig.~\ref{fig::convergence_pillar_opt} to illustrate that the algorithm almost
immediately reaches the local maximum of the objective function, long before the
changeover to the binarization step $q=1$ at the iteration 40 (first dashed
vertical red line). The target efficiency $R_{0,-1}$ oscillates around a constant
value after $q=2$, without dependence on the increasing binarization. This is
why the $q$ binarization parameter changeover is imposed only every 10
iterations (other dashed vertical red lines).

The optimized pattern obtained is shown on 2 main periods (along the $y$ axis)
and 10 periods along the grating lines ($x$ axis) in
Fig.~\ref{fig::pillar_performance}a (3D view) and
Fig.~\ref{fig::pillar_performance}b (view from above) with the associated
spectral response in blue line in Fig.~\ref{fig::pillar_performance}c.
In this figure, we also show the performance obtained with a mesh-based grating
optimized in the conical framework with the same specifications (see
Ref.~\cite{ownArticle} for the full description of the method). While more
freeform, the mesh-based design does not show much better performance than the
pillar-based pattern, demonstrating that in this case, the latter is close to
the maximal possible performance. The primary advantage of this
pillar-based pattern is the flatness of its response above 50\% on both octaves
without forbidden wavelengths, which is preferable even if the average
efficiency is now 57\%. Furthermore, the polarization dependence is restrained,
with the average efficiency for $s$-polarization $\psi_i=90^{\circ}$ at 50\%
(Fig.~\ref{fig::pillar_performance}d). The whole spectrum is reflected with an
efficiency between 35 and 75\%, 80\% of the wavelengths being concentrated in a
20\%-interval ([45--65]\% for $p$-polarization and [35--55]\% for
$s$-polarization). Compared to approaches requiring two distinct
gratings to handle both polarizations~\cite{Sell_Fan_multiwavelength}, this is a
significant step forward. Furthermore, unlike the high-efficiency gratings
reported in Ref.~\cite{nanoblazed_IOF}, the proposed design offers much better
flatness in its response.

This grating satisfies a large number of performance and manufacturability
criteria, which makes it the best candidate for the next blazed metasurface
device to be fabricated. Allowing for a blazing effect on such a wide band
inevitably forces the optimal pattern to avoid any material in the right
zone of the period along the $y$ axis, \textit{i.e.} a large hole of a bit less
than 1\,$\mu$m. This remains a sensitive technical aspect for the fabrication,
but it is not impossible.

In this final section, we have linked the manufacturing capabilities of e-beam
lithography and RIE for nanostructures with our optimization code, providing a
broadband blazed metasurface on two octaves whose fabrication is ongoing.
Starting from non-realistic, freeform structures in the previous section, we
have managed to modify the geometry specifications to obtain a manufacturable
design, while this pillar-based optimization algorithm itself incorporates 
no explicit physical (fabrication) constraints. The flatness of the response of this design on the $-1$st
diffraction order, as well as its limited dependence on polarization are
additional qualities to its global performance, making it a serious candidate in
the landscape of future diffraction devices.

\section{Conclusion}\label{sec::conclusion}

In this work, we present a topology optimization framework for the full 3D
modeling of blazed metasurfaces, enabling the rigorous treatment of
nanostructured broadband gratings in reflection. The scattered-field formulation
of Maxwell’s equations, discretized with the Finite Element Method and combined
with an adjoint-based sensitivity analysis, allows for the efficient
optimization of a large number of degrees of freedom within a fixed design
region.

A first implementation, referred to as mesh-based, directly associates the
density variable to the 3D Finite Element mesh of the design region. This
approach fully exploits the versatility of topology optimization and leads to
freeform patterns exhibiting outstanding broadband performances, with an average
diffraction efficiency of 62\% on order $-$1 over the [400--${1500}$]\,nm range
for the considered incidence. It highlights the potential of metasurfaces to
outperform conventional sawtooth blazed gratings over two octaves. Nevertheless,
the resulting geometries in 3D are computationally very demanding, show no
significant improvement over conical simulations from our previous work, and
remain hardly compatible with realistic nanofabrication processes.

To address this limitation, we introduced a second strategy, denoted
pillar-based, in which the material distribution is constrained to elementary
dielectric pillars compatible with electron-beam lithography and Reactive Ion
Etching. The comparison between mesh-based and pillar-based optimizations
underlines a central result of this study: while unconstrained 3D topology
optimization reveals the ultimate performance limits of broadband blazed
metasurfaces, embedding fabrication constraints at the design stage enables a
realistic and technologically viable trade-off between efficiency and
manufacturability. The obtained metasurface preserves the blazing behavior over
a broad spectral range while remaining fully manufacturable. An average
efficiency of 57\% over two octaves is achieved in $p$-polarization, with a
moderate polarization dependence and a homogeneous spectral response. This 3D
framework therefore establishes a coherent pathway from numerical freeform
design to fabricable nanostructured metasurfaces for next-generation
spectrographs. A quantitative forward error analysis of this design, assessing the
degradation of its efficiency under realistic fabrication uncertainties, is left
for future work and will accompany the experimental characterization of the
sample currently under fabrication.

In conclusion, this paper combines Finite Element modeling, topology
optimization, and nanofabrication, bridging the gap between theoretical design
and real-world manufacturing techniques. Leveraging the concept and tools
developed in this work, we can envision new optical functions with specific
metasurface designs. Beyond the application to spectroscopy, this framework
stands as a versatile tool for the design of advanced metasurfaces. This work is
part of the improvements that can be brought in various domains of physics and
biology, as soon as material characterization is needed. Even though our
motivation originates from the need to increase the throughput of spectrographs
for astronomy applications, it appears that our open-source framework for 
topology optimization can be adapted and applied in many contexts in nanophotonics.

\section{Acknowledgements}
This work has been partially supported by CNES and
Thales Alenia Space with a PhD grant.

The authors would like to thank Roland Salut,
Nicolas Passilly, Andrei Mursa and Quentin Tanguy at FEMTO-ST, Besançon
(France) for fruitful discussions about the nano-patterning constraints of
metasurfaces. The authors would also like to thank Julien Lumeau and Antonin
Moreau at Institut Fresnel, Marseille (France) for fruitful discussions
about the deposition of the layers of the proposed grating.

\bibliographystyle{unsrt}
\footnotesize\bibliography{bibliography}

\normalsize\section*{\hypertarget{app::appendix}{Appendix}}

\begin{appendices}
    \section{Preliminary manufacturability constraints: connectedness and binarization}\label{app::constraints}

    \subsection{Definition of the filters}
    
    Two filters are introduced for this optimization problem. They are necessary yet
    generally not sufficient to generate a manufacturable grating.
    
    \begin{figure}[htb]
        \centering  
        \includegraphics[width=1\textwidth]{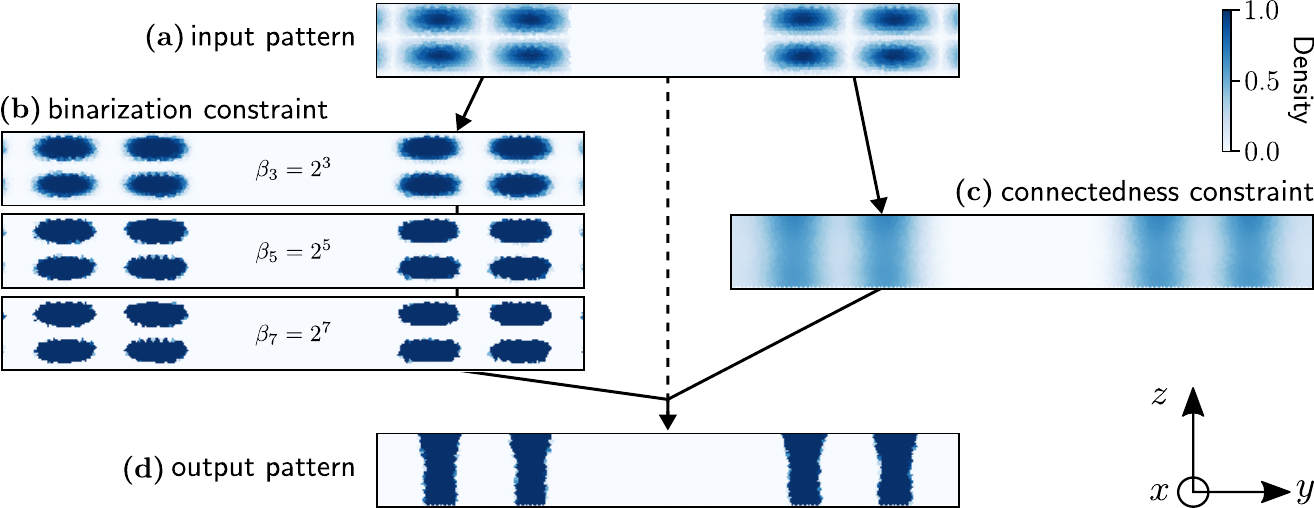}
        \caption{Description of the two filters applied on the density field during
        the optimization. Illustration published in Ref.~\cite{ownProc_SPIE}.
        \textbf{(a)}~2D input distribution with blurred rods spaced with narrow
        trenches. \textbf{(b)}~The binarization filter coerces the field to take
        either the value 1 or 0, and is increasing during the optimization process.
        \textbf{(c)}~The connectedness filter ensures that there are no groups that
        are too small, given a certain threshold. \textbf{(d)}~Effect of the
        composition of both filter.}
        \label{fig::explanation_bin-con}
    \end{figure}
    
    \paragraph{Connectedness.} A connectedness (or diffusion) filter is imposed using
    a connectedness per tetrahedron map $\boldsymbol{\rho}_f$ defined hereafter. It
    is a linear function that averages the values around every tetrahedron with a
    given radius $r_f$. If $\mathcal{T}_i$ is the tetrahedron with the density
    $\rho_i$, let $\text{dist}_{j,r_f}$ be a proximity detection function such that
    \begin{equation}\label{eq::def_dist_rf}
        \text{dist}_{j,r_f}(\rho_i) =
        \left\{
            \begin{array}{llc}
                \text{dist}(\mathcal{T}_i,\mathcal{T}_j) & \mbox{if the latter is smaller than $r_f$}\\
                0                                        & \mbox{else}
            \end{array}
        \right.,
    \end{equation}
    where the distance between two tetrahedra is the distance between their
    barycenters. This distance function selects the surroundings tetrahedra closer
    than the filter radius. Then the diffusion filter function is defined by
    Ref.~\cite{TD_3D_TopOpt}: 
    \begin{equation}\label{eq::def_diffusion_filter}
        \rho_{f,i}(\boldsymbol{\rho}) = \frac{\sum_{j=1}^N \text{dist}_{j,r_f}(\rho_i) \rho_j}{\sum_{j=1}^N \text{dist}_{j,r_f}(\rho_i)}.
    \end{equation}
    The effect of the connectedness filter is illustrated in
    Fig.~\ref{fig::explanation_bin-con}c for an input pattern given in
    Fig.~\ref{fig::explanation_bin-con}a. Essentially, small gaps are filled by a
    spatial sliding average function with given radius~$r_f$.
    
    \paragraph{Binarization.} The binarization constraint is applied
    over the connectedness filter. It is given by the usual function~\cite{ADCode}
    denoted by $\hat{\boldsymbol{\rho}}$:
    \begin{equation}\label{eq::filter}
        \hat{\boldsymbol{\rho}}: [0,1]^N \ni \boldsymbol{\rho}_f
        \mapsto \frac{\tanh(\beta_f\nu) + \tanh[\beta_f(\boldsymbol{\rho}_f - \nu)]}
               {\tanh(\beta_f\nu) + \tanh[\beta_f(1 - \nu)]} \in [0,1]^{M_d}
    \end{equation}
    with $\nu\in[0,1]$ and where $\beta_f$ monitors the binarization strength which
    is increased during the optimization process. Indeed, it is crucial to obtain a
    binary design only made of material~1 with permittivity $\varepsilon_{r,1}$ and
    material~2 with permittivity $\varepsilon_{r,2}$, avoiding intermediate values
    of permittivity that would not correspond to any existing optical bulk material.
    The threshold parameter $\nu$, as indicated by its name, defines for which value
    of $\rho_i$ the density is rather lowered down to 0 ($\rho_i \leq \nu$) or
    pushed up to 1 ($\rho_i > \nu$). A standard configuration is $\nu = 1/2$ and
    $\beta_{f,q} = 2^q$ with $q = 1,\dots,7$ increasing during the optimization
    process. The effect of the binarization filter is illustrated in
    Fig.~\ref{fig::explanation_bin-con}b for various values of $\beta_{f,q}$ and for
    an input pattern given in Fig.~\ref{fig::explanation_bin-con}a.
    
    \paragraph{Composition of filters.} To summarize, the constraints on the design
    variables are gathered in one function that coerces the design into having a
    specific behavior. It can be written through a composition of maps:
    \begin{equation}\label{eq::comp_epsilon_appendix}
        \hat{\boldsymbol{\rho}}_f: [0,1]^{M_d} \ni \boldsymbol{\rho} \mapsto (\hat{\boldsymbol{\rho}} \circ \boldsymbol{\rho}_f)(\boldsymbol{\rho}) \in [0,1]^{M_d}.
    \end{equation}
    The effect of applying both the connectedness and binarization filters is
    illustrated in Fig.~\ref{fig::explanation_bin-con}d for the input pattern shown
    in Fig.~\ref{fig::explanation_bin-con}a. Here, the output is 4 vertical pillars,
    where the small initial trenches have been erased.
    
    \subsection{Analytical derivatives of the filters}
    
    \paragraph{Connectedness.} Let $\mathbb{D}$ be the matrix of the different values
    of $\text{dist}_{j,r_f}$: $\mathbb{D}_{i,j} = \text{dist}_{j,r_f}(\rho_i)$ and
    let $\mathds{1}$ be the column vector with ones. Then we can rewrite the
    filter function
    \begin{equation}\label{eq::def_diffusion_filter_mat}
        \rho_{f,i}(\boldsymbol{\rho}) = \frac{\mathbb{D}\boldsymbol{\rho}}{\mathbb{D}\mathds{1}}
    \end{equation}
    where the division of vectors in the last expression designates the division
    element-wise. This expression directly leads to the Jacobian of the filter:
    \begin{equation}\label{eq::jac_diffusion_filter_mat}
        \frac{\partial \boldsymbol{\rho}_f}{\partial \boldsymbol{\rho}}(\boldsymbol{\rho}) = \frac{\mathbb{D}}{\mathbb{D}\mathds{1}}
    \end{equation}
    where the division of a matrix by a vector in the last expression is the
    division of each line of the matrix by this vector (still element-wise).
    
    \paragraph{Binarization.} A straightforward analytical differentiation
    leads to
    \begin{equation}\label{eq::derivative_binarization}
        \frac{\partial \hat{\rho}_i}{\partial \rho_{f,j}}(\boldsymbol{\rho}_f)
        = \frac{\beta_f (1-\tanh^2[\beta_f(\rho_{f,i} - \nu)])}{\tanh(\beta_f\nu) + \tanh[\beta_f(1 - \nu)]} \delta_{ij},
    \end{equation}
    where $\delta_{ij}$ is the Kronecker delta. The matrix multiplication
    \eqref{eq::jac_diffusion_filter_mat}$ \times
    $\eqref{eq::derivative_binarization} (in this order) provides the filtering
    factor appearing in the derivative of $\mathcal{F}_{n,m}$.
    
    \section{Comparison between the conical and the 3D mesh-based optimization processes}\label{app::comparison}
    
    The 3D mesh-based optimized pattern achieved with our solver shows outstanding performance for blazed gratings. However we have observed that running the optimization with the same parameters (angles of incidence, targeted wavelengths, material, etc.) with our conical solver\footnote{\url{https://gitlab.onelab.info/doc/models/-/tree/master/DiffractionGratingsTopOpt/meshGrating_conical_topopt}} yields a very close optimized pattern, regarding its shape and response. The latter is plotted in the paper (see Fig.~5) but for the sake of clarity, we did not display the pattern itself. We only mentioned that it is close to the conical one. This supplemental document provides additional results to this part.
    
\begin{figure}[htb]
        \centering
        \includegraphics[width=.6\textwidth]{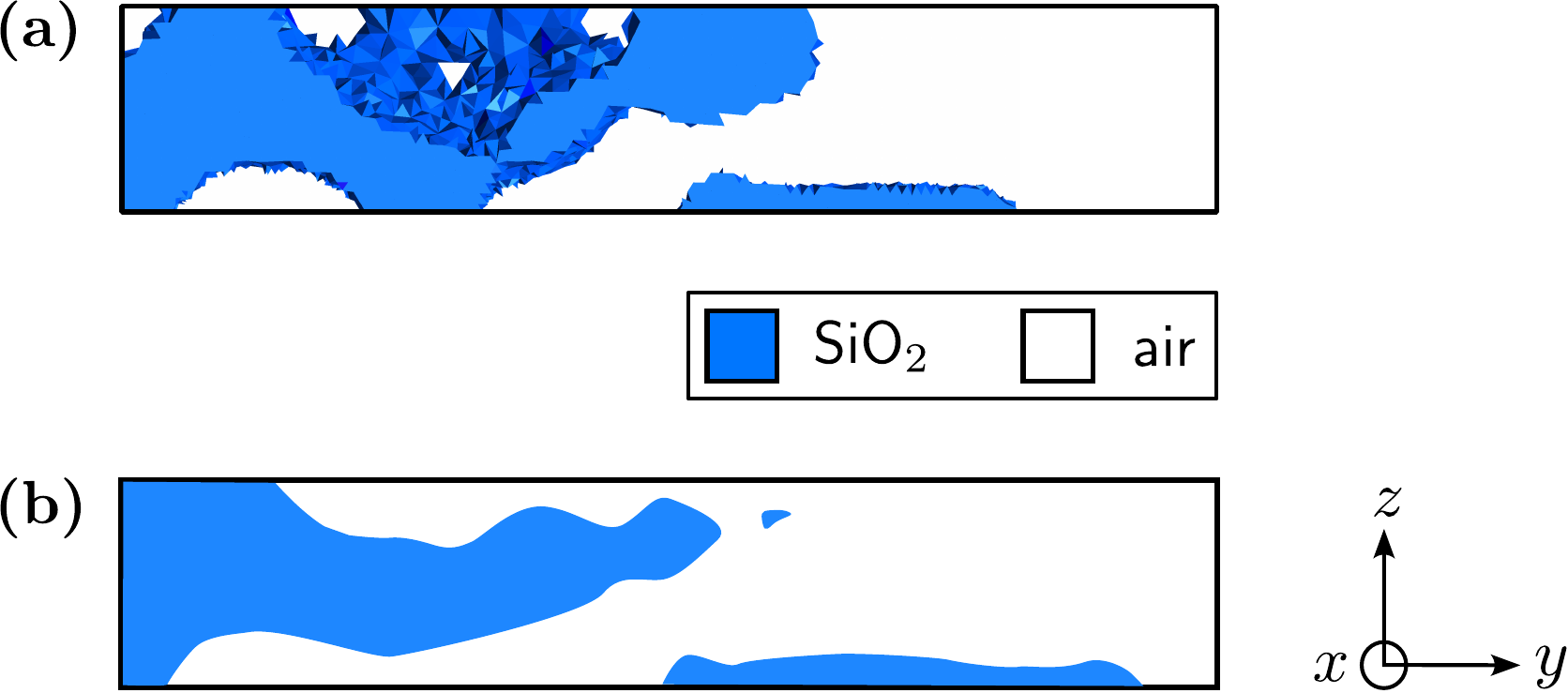}
        \caption{\textbf{(a)}~Side view of the 3D optimized pattern with the set of 17 targeted wavelengths displayed in the paper (Fig.~5). \textbf{(b)}~Achieved pattern with the exact same optimization, but with a conical solver. The boundaries have been smoothed, however the mesh is twice as fine as in 3D.}
        \label{fig::comparison_conical_3D}
    \end{figure}

    Figure~\ref{fig::comparison_conical_3D}a shows a side view of the 3D pattern to illustrate that the obtained shape closely corresponds to that of the output of the same optimization with a conical solver, meaning the mono-periodic grating shown in Fig.~\ref{fig::comparison_conical_3D}b. The degree of freedom along the $x$ axis is harnessed though, with a variation of shape between $x_{\min} = 0$ and $x_{\max} = 330$\,nm.
    
    \begin{figure}[htb]
        \centering
        \includegraphics[width=.6\textwidth]{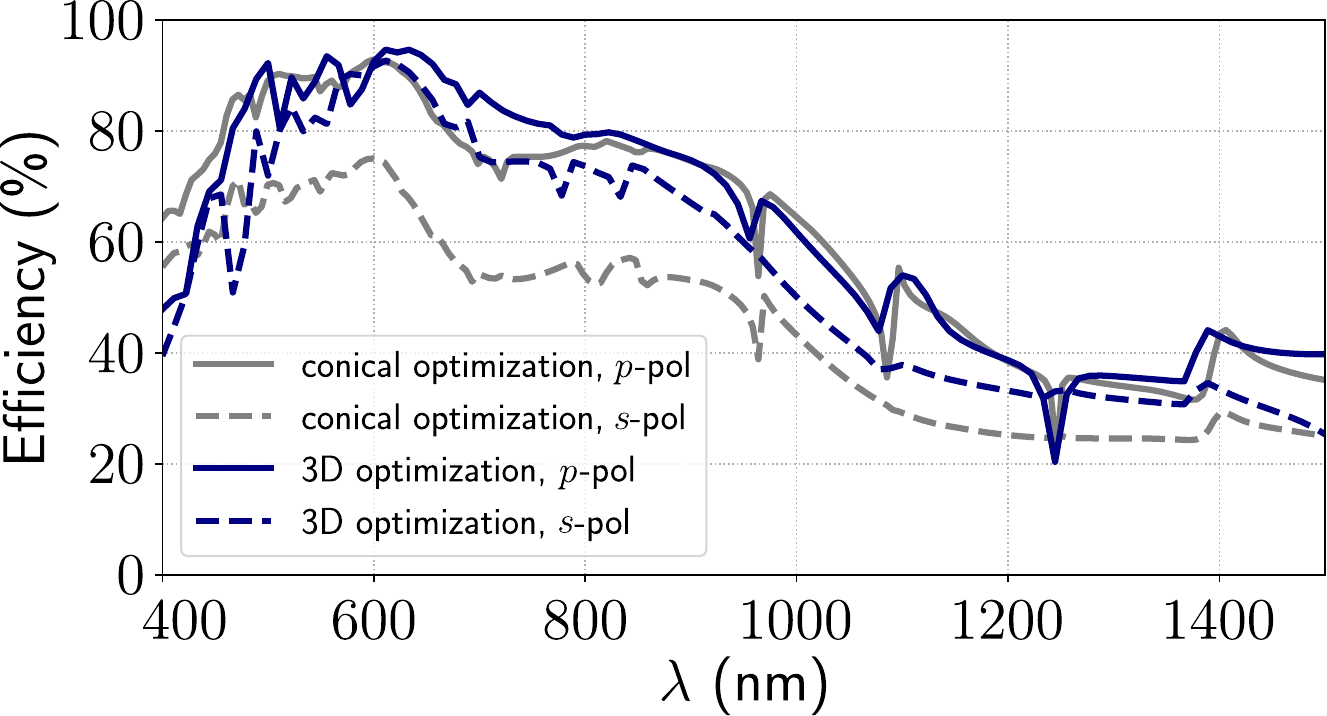}
        \caption{Sepctral response of the bi-periodic grating of Fig.~\ref{fig::comparison_conical_3D}a (blue lines) and the mono-periodic grating of Fig.~\ref{fig::comparison_conical_3D}b (gray lines), for $p$-polarization $\psi_i = 0^{\circ}$ (solid lines) and $s$-polarization $\psi_i = 90^{\circ}$ (dashed lines).}
        \label{fig::comparison_conical_3D_polarization}
    \end{figure}
    
    Unfortunately, this variation of shape does not improve the overall blazing effect, as shown by the plot of Fig.~5c (blue vs. gray curve). However, the polarization dependence is much lower with the 3D pattern, as illustrated in Fig.~\ref{fig::comparison_conical_3D_polarization}. In this graph, the $s$-polarization $\psi_i = 90^{\circ}$ (dashed lines) is added to the $p$-polarization from Fig.~5c of the article (solid lines).
    
    In conclusion, the bi-periodic pattern is more satisfactory, for it is mechanically sustainable and shows low polarization dependence, but it takes 10 days to run this optimization (vs. 9 hours for a conical study), yielding a result that still shows challenging shapes for manufacturing.
\end{appendices}

\end{document}